\documentclass[aps,prl,twocolumn,floatfix,superscriptaddress,nofootinbib,showpacs,showkeys,preprintnumbers]{revtex4}
\usepackage{graphicx,amsmath}

\providecommand{\Journal}[4] {#1 {\bf #2}, #3 (#4)}
\providecommand{\CPC}{Comput. Phys. Commun. } %
\providecommand{\MPLA}{Mod. Phys. Lett. A} %
\providecommand{\NIMA}{Nucl. Instr. Meth. A } %
\providecommand{\NPA}{Nucl. Phys. A } %
\providecommand{\NPB}{Nucl. Phys. B } %

\providecommand{\PAN}{Phys. Atom. Nucl. } %
\providecommand{\PLB}{Phys. Lett. B } %
\providecommand{\PRL}{Phys. Rev. Lett. } %
\providecommand{\PRC}{Phys. Rev. C } %
\providecommand{\PRD}{Phys. Rev. D } %
\providecommand{\YF}{Yad. Fiz. } %
\providecommand{\ZPA}{Z. Phys. A } %
\newcommand{\pythia}{{\sc Pythia6}}
\newcommand{\nsts}{{N_s^{2\sigma}}}
\newcommand{\nbts}{{N_b^{2\sigma}}}

\begin{document}

\title{Evidence for a narrow $\mathbf{|S|=1}$ baryon state at a mass of 1528\,MeV in
quasi-real photoproduction \\
 }


\def\groupalberta{\affiliation{Department of Physics, University of Alberta, Edmonton, Alberta T6G 2J1, Canada}}
\def\groupargonne{\affiliation{Physics Division, Argonne National Laboratory, Argonne, Illinois 60439-4843, USA}}
\def\groupbari{\affiliation{Istituto Nazionale di Fisica Nucleare, Sezione di Bari, 70124 Bari, Italy}}
\def\groupbeijing{\affiliation{School of Physics, Peking University, Beijing 100871, China}}
\def\groupchina{\affiliation{Department of Modern Physics, University of Science and Technology of China, Hefei, Anhui 230026, China}}
\def\groupcolorado{\affiliation{Nuclear Physics Laboratory, University of Colorado, Boulder, Colorado 80309-0446, USA}}
\def\groupdesy{\affiliation{DESY, Deutsches Elektronen-Synchrotron, 22603 Hamburg, Germany}}
\def\groupzeuthen{\affiliation{DESY Zeuthen, 15738 Zeuthen, Germany}}
\def\groupdubna{\affiliation{Joint Institute for Nuclear Research, 141980 Dubna, Russia}}
\def\grouperlangen{\affiliation{Physikalisches Institut, Universit\"at Erlangen-N\"urnberg, 91058 Erlangen, Germany}}
\def\groupferrara{\affiliation{Istituto Nazionale di Fisica Nucleare, Sezione di Ferrara and Dipartimento di Fisica, Universit\`a di Ferrara, 44100 Ferrara, Italy}}
\def\groupfrascati{\affiliation{Istituto Nazionale di Fisica Nucleare, Laboratori Nazionali di Frascati, 00044 Frascati, Italy}}
\def\groupgent{\affiliation{Department of Subatomic and Radiation Physics, University of Gent, 9000 Gent, Belgium}}
\def\groupgiessen{\affiliation{Physikalisches Institut, Universit\"at Gie{\ss}en, 35392 Gie{\ss}en, Germany}}
\def\groupglasgow{\affiliation{Department of Physics and Astronomy, University of Glasgow, Glasgow G12 8QQ, United Kingdom}}
\def\groupillinois{\affiliation{Department of Physics, University of Illinois, Urbana, Illinois 61801-3080, USA}}
\def\groupmit{\affiliation{Laboratory for Nuclear Science, Massachusetts Institute of Technology, Cambridge, Massachusetts 02139, USA}}
\def\groupmichigan{\affiliation{Randall Laboratory of Physics, University of Michigan, Ann Arbor, Michigan 48109-1120, USA }}
\def\groupmoscow{\affiliation{Lebedev Physical Institute, 117924 Moscow, Russia}}
\def\groupmunich{\affiliation{Sektion Physik, Universit\"at M\"unchen, 85748 Garching, Germany}}
\def\groupnikhef{\affiliation{Nationaal Instituut voor Kernfysica en Hoge-Energiefysica (NIKHEF), 1009 DB Amsterdam, The Netherlands}}
\def\groupstpetersburg{\affiliation{Petersburg Nuclear Physics Institute, St. Petersburg, Gatchina, 188350 Russia}}
\def\groupprotvino{\affiliation{Institute for High Energy Physics, Protvino, Moscow region, 142281 Russia}}
\def\groupregensburg{\affiliation{Institut f\"ur Theoretische Physik, Universit\"at Regensburg, 93040 Regensburg, Germany}}
\def\grouprome{\affiliation{Istituto Nazionale di Fisica Nucleare, Sezione Roma 1, Gruppo Sanit\`a and Physics Laboratory, Istituto Superiore di Sanit\`a, 00161 Roma, Italy}}
\def\groupsimonfraser{\affiliation{Department of Physics, Simon Fraser University, Burnaby, British Columbia V5A 1S6, Canada}}
\def\grouptriumf{\affiliation{TRIUMF, Vancouver, British Columbia V6T 2A3, Canada}}
\def\grouptokyo{\affiliation{Department of Physics, Tokyo Institute of Technology, Tokyo 152, Japan}}
\def\groupamsterdam{\affiliation{Department of Physics and Astronomy, Vrije Universiteit, 1081 HV Amsterdam, The Netherlands}}
\def\groupwarsaw{\affiliation{Andrzej Soltan Institute for Nuclear Studies, 00-689 Warsaw, Poland}}
\def\groupyerevan{\affiliation{Yerevan Physics Institute, 375036 Yerevan, Armenia}}


\groupalberta
\groupargonne
\groupbari
\groupbeijing
\groupchina
\groupcolorado
\groupdesy
\groupzeuthen
\groupdubna
\grouperlangen
\groupferrara
\groupfrascati
\groupgent
\groupgiessen
\groupglasgow
\groupillinois
\groupmit
\groupmichigan
\groupmoscow
\groupmunich
\groupnikhef
\groupstpetersburg
\groupprotvino
\groupregensburg
\grouprome
\groupsimonfraser
\grouptriumf
\grouptokyo
\groupamsterdam
\groupwarsaw
\groupyerevan


\author{A.~Airapetian}  \groupyerevan
\author{N.~Akopov}  \groupyerevan
\author{Z.~Akopov}  \groupyerevan
\author{M.~Amarian}  \groupzeuthen \groupyerevan
\author{V.V.~Ammosov}  \groupprotvino
\author{A.~Andrus}  \groupillinois
\author{E.C.~Aschenauer}  \groupzeuthen
\author{W.~Augustyniak}  \groupwarsaw
\author{R.~Avakian}  \groupyerevan
\author{A.~Avetissian}  \groupyerevan
% ...

\collaboration{The HERMES Collaboration} \noaffiliation


\begin{abstract}

Evidence for a narrow baryon state is found in quasi-real
photoproduction on a deuterium target through the decay channel $p
K^0_S \to p \pi^+ \pi^-$. A peak is observed in the $p K^0_S $
~invariant mass spectrum at $1528 \pm 2.6\mbox{(stat)} \pm
2.1\mbox{(syst)}$\,MeV.  Depending on the background model,
the na{\"i}ve statistical significance of the peak is 4--6
standard deviations and its width may be somewhat larger than
the experimental resolution of $\sigma=4.3$--6.2\,MeV.
This state may be interpreted as the predicted $S$=$+1$ exotic
$\Theta^{+}(uudd\bar{s})$ pentaquark baryon. No signal for an
hypothetical $\Theta^{++}$ baryon was observed in the $pK^+$
invariant mass distribution.  The absence of such a signal
indicates that an isotensor $\Theta$ is excluded and an isovector
$\Theta$ is unlikely.

\end{abstract}



\maketitle

One of the central mysteries of hadronic physics has been the
failure to observe baryon states beyond those whose
quantum numbers can be explained in terms of three quark
configurations. Exotic hadrons with manifestly more complex quark
structures, in particular
exotics consisting of five quarks, were proposed on the basis
of quark and bag models~\cite{Jaff76} in the early days of QCD.
The hope has been that the discovery of such objects would
provide new insights into the dynamics of quark interactions in
the strong coupling regime.  Although it was hypothesized~\cite{Lip87}
that pentaquark systems involving heavy quarks, e.g.
$uud\bar{c}s$, offered the most promising prospects for isolating
such exotics, experimental searches carried out at
FNAL~\cite{Moi96} found no evidence for such states.

From quite a different point of view,
it was noted~\cite{Man84,Che85} that the Skyrme model predicts new exotic
states belonging to higher SU(3) representations. Using this model,
Praszalowicz~\cite{Pra03} provided the first estimate of the mass of
the lightest exotic state, $M$$\approx$1530\,MeV. Subsequently, an
exotic baryon of spin 1/2, isospin 0, and strangeness $S$=$+1$ was
discussed as a feature of the Chiral Quark Soliton model~\cite{Che85}.
In this approach~\cite{Wal92,Dia97} the baryons are rotational states
of the soliton nucleon in spin and isospin space, and the lightest
exotic baryon lies at the apex of an anti-decuplet with spin 1/2,
which corresponds to the third rotational excitation in a three flavor
system. Treating the known N(1710) resonance as a member of the
anti-decuplet, Diakanov, Petrov, and Polyakov~\cite{Dia97} derived a
mass of 1530\,MeV and a width of less than 15\,MeV for this exotic
baryon, since named the $\Theta^+$. It corresponds to a
$uudd\overline{s}$ configuration, and decays through the channels
$\Theta^{+}\rightarrow pK^0$ or $nK^{+}$. However, measurements of
$K^+$ scattering on proton and deuteron targets showed no
evidence~\cite{Arn03} for strange baryon resonances, and appear to
limit the width to remarkably small values of order an MeV.
Presumably, the difficulty of experiments with kaon beams so close to
threshold severely limited the sensitivity to such narrow excitations.
In a recent review of this subject~\cite{GBQ99}, an experimental
search by means of electro-production was suggested.

Experimental evidence for an exotic baryon first came
recently~\cite{SPring8} from the observation of a narrow
resonance at $1540\pm 10\mbox{(syst)}$\,MeV in the $K^-$ missing mass spectrum
for the $\gamma n \to K^+K^-n$ reaction on $^{12}$C. The decay
mode corresponds to a $S$=$+1$ resonance and signals an exotic
pentaquark state with quark content ($uudd\bar{s}$). Confirmation
came quickly from a series of experiments, with the observation
of sharp peaks~\cite{DIANA,JLab,SAPHIR,Asr03,JLabp} in the $nK^+$ and
$pK^0_S$ invariant mass spectra near 1540\,MeV, in each case with
a width limited by the experimental resolution. The failure to
observe a corresponding $\Theta^{++}$ peak in the $pK^+$
invariant mass spectrum in some of these
experiments~\cite{SAPHIR,JLab} was taken to suggest that the state is an
isospin singlet.

Alternative theoretical explanations have been proposed recently
to explain this new exotic state.  In one model, the $\Theta$ is
described as an isotensor pentaquark~\cite{Cap03}, so that the
narrow width results from the isospin-violating strong decay.
A search for the decay of the isospin partners such as the
$\Theta^{++} (uuud\bar{s})$ can provide a strong test of this
idea.  In a second interpretation, Karliner and Lipkin have
developed a cluster model using a diquark-triquark
configuration~\cite{KarLip}, in which the $\Theta^+$ is also a
positive-parity isosinglet member of an antidecuplet.  Thirdly,
using a model based on the  strong color-spin correlation force,
Jaffe and  Wilczek~\cite{JW03} propose that the $\Theta^+$
consists of two  highly correlated $ud$ pairs coupled to an
$\bar{s}$.  In their picture the positive-parity isosinglet 
$\Theta^+$ lies at the apex of
an nearly-ideally  mixed SU(3)$_f$ $\mathbf{\mathbf{8_f}} \oplus
\overline{\mathbf{10_f}}$ multiplet.  The narrowness of the
1530\,MeV state is attributed to the relatively weak coupling of
the kaon-nucleon continuum to the pentaquark $[ud]^2\bar{s}$
configuration.

The baryon states at the bottom two vertices of the anti-decuplet
must also be manifestly exotic.  Strong evidence in support of the
baryon decuplet comes from the reported observation of an exotic
$S$=$-2$, $Q$=$-2$ baryon resonance in proton-proton collisions
at $\sqrt{s}=17.2$\,GeV at the CERN SPS~\cite{Alt03}. A narrow
peak at a mass of about $1862$\,MeV in the $\Xi^-\pi^-$ invariant
mass spectrum is proposed as a candidate for the predicted exotic
$\Xi^{--}_{\frac{3}{2}}$ baryon with $S$=$-2$,
$I$=${\frac{3}{2}}$ and a quark content of ($dsds\bar{u}$). At
the same mass, a peak is observed that is a candidate for the
$\Xi^{0}_{\frac{3}{2}}$ member of this isospin quartet. The
corresponding anti-baryon spectra show enhancements at the same
invariant mass.
This observed mass of $1862$\,MeV falls between the predictions of
Refs.~\cite{Dia97} and \cite{JW03}, although closer to the latter.
Also, the positive parity for the $\Theta^+$ predicted by
these models contrasts with the negative parity suggested by the
first lattice results~\cite{Sas03,Csi03}.
The general theoretical situation is still quite unsettled.

This Letter presents the results of a search for the $\Theta^+$ in
quasi-real photoproduction  on deuterium. In addition to corroborating
some features of the state measured previously, the data reported here
provide more restrictive information related to its mass and isospin. The
data were obtained by the HERMES experiment with the 27.6\,GeV
positron beam of the HERA storage ring at DESY. Stored beam currents
ranged from 9 to 45\,mA. An integrated luminosity of 250\,pb$^{-1}$
was collected on a longitudinally polarized deuterium gas target. The
yields were summed over two spin orientations.

The HERMES spectrometer~\cite{SPE} consists of two identical
halves located above and below the positron beam pipe, and has an
angular acceptance of $\pm 170$\,mrad horizontally, and $\pm$(40
-- 140)\,mrad vertically. The trigger was formed by either a
coincidence between scintillating hodoscopes, a preshower
detector and a lead-glass calorimeter, or a coincidence between
three scintillating hodoscopes and two tracking planes, requiring
that at least one charged track appears in each of the detector
halves of the spectrometer.

The analysis searched for inclusive photoproduction of the
$\Theta^+$ followed by the decay $\Theta^+\to p K^0_S \to p \pi^+
\pi^-$. Events selected contained at least three tracks: two
oppositely charged pions in coincidence with one proton.
Identification of charged pions and protons was accomplished with
a Ring-Imaging \v{C}erenkov (RICH) detector~\cite{RICH} which
provides separation of pions, kaons and protons over most of the
kinematic acceptance of the spectrometer. The RICH identification
efficiencies and cross contaminations had been determined in a
limited kinematic domain using known particle species from
identified resonance decays. However, because the RICH
performance is sensitive to event topology, it was essential to
determine these efficiencies and contaminations for pions and
protons under conditions as close as possible to those of the
present measurement. This was accomplished by means of a Monte
Carlo simulation based on the \pythia\ generator discussed below.
Events with the relevant topology were used to determine these
parameters as a function of particle momentum. The data from the
simulation indicated that cross contaminations would be
negligible if protons were restricted to a momentum range of
4--9\,GeV/c and pions to a range of 1--15\,GeV/c, the kinematic
restrictions subsequently used in the analysis.\\

\begin{figure} [htb]
\begin{center}
\includegraphics[width=7.0cm,angle=0]{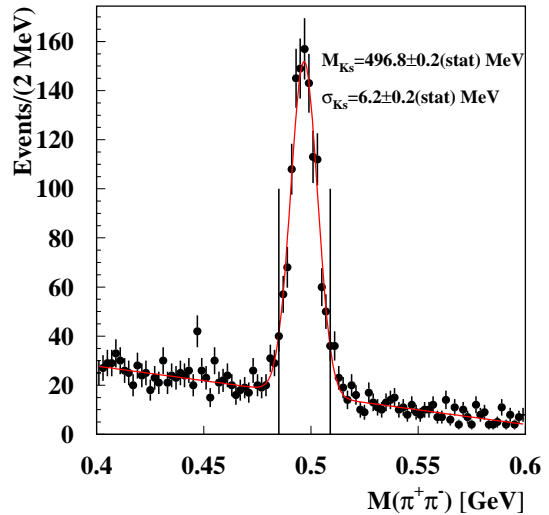}
\end{center}
\vspace{-5mm}

\caption{Invariant mass distribution of two oppositely charged pions,
subject to the constraints in event topology discussed in the text. A
window corresponding to $\pm 2\,\sigma$ is shown by the vertical
lines.}

\label{fig1}
\end{figure}

The event selection included constraints on the event topology to
maximize the yield of the $K^0_S$ peak in the $M_{\pi^+\pi^-}$
spectrum while minimizing its background. However, no constraints
were optimized to increase the significance of the signal visible
in the final $M_{p \pi^+\pi^-}$ spectrum, as such optimization
would have produced a spectrum to which standard statistical
tests do not apply. Based on the intrinsic tracking
resolution, the required event topology included a minimum distance of
approach between the two pion tracks less than 1\,cm (the midpoint of which
is defined as the $K^0_S$ decay vertex), a minimum distance of approach
between the proton and reconstructed $K^0_S$ tracks less than 6\,mm (the
midpoint of which is defined as the production vertex), a
radial distance of the production vertex from the positron beam
axis less than 4\,mm, a $z$ coordinate of the production vertex
within the $\pm 20$\,cm long target cell of $-18$\,cm$ < z < +18\,$cm
along the beam direction, and a $K^0_S$ decay
length (separation of production and $K^0_S$ decay vertices) greater than
7\,cm.  To suppress contamination from the $\Lambda(1116)$ hyperon,
events were rejected where the invariant mass $M_{p \pi^-}$ fell
within $2\,\sigma$ of the nominal $\Lambda$ mass, where $\sigma=2.6$\,MeV
is the apparent width of the $\Lambda$ peak observed in this experiment.

The resulting invariant $M_{\pi^+\pi^-}$ spectrum is shown in Fig.~1.
The position of the $K^0_S$ peak is within 1\,MeV of the expected
value of $497.7 \pm 0.03$\,MeV~\cite{PDG}. To search for the
$\Theta^+$, events were selected with a $M_{\pi^+\pi^-}$ invariant
mass within $\pm 2\,\sigma$ about the centroid of the $K_S^0$ peak.
The resulting spectrum of the invariant mass of the $p\pi^+\pi^-$
system is displayed in Fig.~2. A narrow peak is observed.
There is no known positively charged strangeness-containing baryon in
this mass region (other than the $\Theta^+$) that could account for
the  observed peak.  Also, the $M_{p\pi^+\pi^-}$ spectrum corresponding
to the sideband background adjacent to the $K^0_S$ peak in Fig.~1
was found to be featureless.

\begin{figure} [htb]
\begin{center}
\includegraphics[width=7.0cm,angle=0]{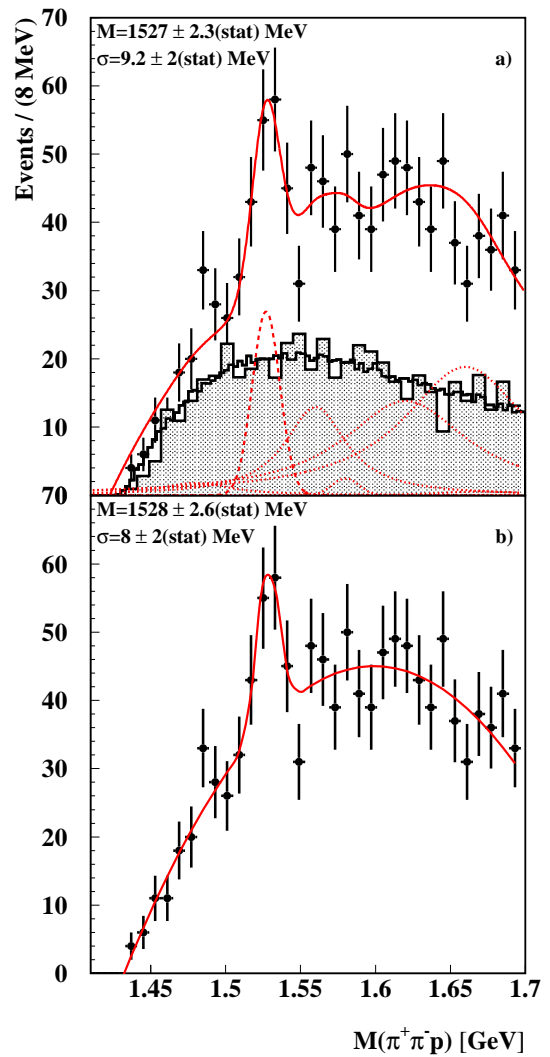}
\end{center}

\caption{Distribution in invariant mass of the $p \pi^+\pi^-$ system
subject to various constraints described in the text. The
experimental data are represented by the filled circles with
statistical error bars, while the
fitted smooth curves result in the indicated position and $\sigma$
width of the peak of interest. In panel a), the \pythia\ Monte Carlo simulation
is represented by the gray shaded histogram, the mixed-event model 
normalised to the \pythia\ simulation is
represented by the fine-binned histogram, and the fitted curve is
described in the text. In panel b), a fit to the data of a Gaussian
plus a third-order polynomial is shown.}
\label{fig2}
\end{figure}

The non-resonant contribution to the spectrum was estimated by means
of a simulation using a version of the \pythia\ code~\cite{PYTHIA6}
tuned for HERMES kinematics~\cite{ourPYTHIA6}. This event generator
contains no resonances in the mass range of Fig.~2a that decay in the
$p K_S^0$ channel. The resulting simulated spectrum is shown in
Fig.~2a as the gray hatched histogram.
The statistical precision of the present study is limited by the rare
topology of the events selected.  Trigger inefficiencies were not
included in the simulation, but are believed to be small.
The simulated spectrum falls below the data at high invariant mass
where $\Sigma^{*+}$ resonances are known to exist~\cite{PDG}.
Therefore, if \pythia\ is assumed to be capable of describing the
shape of the non-resonant contribution, it can be concluded that there
is substantial resonant strength distributed over the high-mass
portion of the spectrum. At the position of the observed peak in the
data, no corresponding structure appears in the simulated spectrum.

\begin{table*}[t]
\caption{Mass values and experimental widths, with their
statistical and  systematic uncertainties, for the
$\Theta^+$ from the two fits, labelled by a) and b), shown in the
corresponding panels of Fig.~2.  Rows a') and b') are based on the same
background models as rows a) and b) respectively,
but a different mass reconstruction expression that
is expected to result in better resolution.  Also shown are the number of
events in the peak $N_s$ and the background $N_b$, both evaluated
from the functions fitted to the mass distribution, and the
results for the na{\"i}ve significance $\nsts/\sqrt{\nbts}$ and realistic
significance $N_{s}/{\delta N_{s}}$.  The systematic uncertainties
are common (correlated) between rows of the table. }
\label{tab1}
\begin{ruledtabular}
\begin{tabular}{lccccccc}
\hspace*{5mm} &
$\Theta^+$ mass & FWHM & $\nsts$& $\nbts$& na{\"i}ve & Total & signif.\\
 & [MeV]& [MeV] & in $\pm2\sigma$ &in $\pm2\sigma$ &
 signif. & $N_s\pm\delta N_s$ \\ \hline
a)  & $1527.0\pm2.3\pm2.1$ & $22\pm5\pm2$ & 74 & 145 & $6.1\,\sigma$
& $78\pm18$ & $4.3\,\sigma$   \\
a')  & $1527.0\pm2.5\pm2.1$ & $24\pm5\pm2$ & 79 & 158 & $6.3\,\sigma$
& $83\pm20$ & $4.2\,\sigma$   \\
b)  & $1528.0\pm2.6\pm2.1$ & $19\pm5\pm2$ & 56 & 144 & $4.7\,\sigma$
& $59\pm16$ & $3.7\,\sigma$ \\
b')  & $1527.8\pm3.0\pm2.1$ & $20\pm5\pm2$ & 52 & 155 & $4.2\,\sigma$
& $54\pm16$ & $3.4\,\sigma$   \\
\end{tabular}
\end{ruledtabular}
\end{table*}
In order to determine the centroid, width and significance of the
peak observed in Fig.~2, three different models for the
background were explored.  For the first model, the \pythia\
simulation is taken to represent the non-resonant background, and
the remaining  strength in the spectrum is attributed to a
combination of known broad resonances and a new structure near
1.53\,GeV.  For the second model, it is assumed that the
non-resonant background involves a large enough typical
multiplicity that the 4-momenta of the $K_S^0$ and proton are
largely uncorrelated. In this case, this background can be
simulated by combining from different events a kaon and proton
that satisfy the same kinematical requirements as the tracks
taken from single events in the main analysis. Since resonances
are typically visible only as rare correlations
between their decay particles, their contributions will be
relatively suppressed in this method. Fig.~2a shows that this
procedure yields a shape that is very similar to that from the
\pythia\ simulation, within the available statistics. By fitting
a polynomial to the mixed-event background normalized to the \pythia\
simulation, and then fitting this polynomial together with the
amplitudes of peaks for six known $\Sigma^{*+}$
resonances in the mass range shown in Fig.~2 (dotted curves),
plus all parameters of a narrow Gaussian (dashed curve) for the peak of
interest, a good description of the entire spectrum is obtained.
This procedure is intended to demonstrate that the background
is consistent with known information.
The included $\Sigma^{*+}$ resonances were assigned fixed values
of $M=1480$\,MeV with $\Gamma=55$\,MeV (PDG status = *),
$M=1560$\,MeV with $\Gamma=47$\,MeV (**), $M=1580$\,MeV with
$\Gamma=13$\,MeV (**), $M=1620$ and 1660\,MeV with $\Gamma=100$\,MeV (***),
and $M=1670$\,MeV with $\Gamma=60$\,MeV (****)~\cite{PDG}.
Each intrinsic Breit-Wigner width was taken
as the midpoint of the range of listed measurements, and was then
augmented by an
instrumental resolution of FWHM$=14.3$\,MeV added in quadrature.
Since the $\Sigma^+(1580)$ has a width smaller than the instrumental
resolution, it was taken to be Gaussian with $\sigma=8.9$\,MeV.
The numerical results of the fit are given in row a) of
Table~\ref{tab1}.  In addition, row a') shows the result of
applying the same method to a mass spectrum based on
an expression that is expected to provide an
instrumental resolution improved by about 30\%:
\begin{eqnarray}
{M'_{pK^0_S}}^2 &\equiv& \Big(\sqrt{M_p^2+\mathbf{p}_p^2}+
\sqrt{M_{K^0_S}^2+\mathbf{p}_{K^0_S}^2}\Big)^2
\nonumber\\ && - (\mathbf{p}_p+\mathbf{p}_{K^0_S})^2\,, \label{eq:massdiff}
\end{eqnarray}
where $M_{K^0_S}$ is taken from the PDG~\cite{PDG}.
The third approach is based on the hypothesis that all of the
background strength in the observed spectrum (apart from the narrow
peak) can be described by a slowly varying function that extends under
the feature of interest. Hence, the spectrum was fit with a Gaussian
plus a polynomial. The appropriate degree of the polynomial used in
the fit was determined by comparing results using orthonormal
Chebyshev polynomials of various degrees. The curve shown in Fig.~2b
results from a fit with a third-order polynomial, and rows~b) and b') of
Table~\ref{tab1} gives the numerical values from so fitting the two spectra
corresponding to rows a) and a').

More specifically, Table~\ref{tab1} compares the results from the two
fits for the centroid of the peak of interest, its width and the
statistical significance according to two different prescriptions
discussed below. The resulting values for the centroid are found to be
consistent, while the width and significance depend
on the method chosen to describe the remaining
strength of the spectrum. Table~\ref{tab1} also lists for both fits the
number of events given by the fitted function for the peak of interest ($\nsts$)
as well as for the background ($\nbts$), in the invariant mass interval
corresponding to $\pm 2\,\sigma$.  The full area $N_s$ of the Gaussian
fitted to the peak of interest is also given with its uncertainty from
the fit.  This area itself, together with the width, were chosen to be
explicit fit parameters to avoid the effect on the uncertainty in the area
of correlations between the amplitude and the width or background parameters.
All of these
results are from unbinned maximum likelihood fits~\cite{RooFit} to the
original event distributions, as it was found that the results of
fitting the histograms shown in Fig.~2 can be sensitive to the choice
of bin size or starting offset.

Several alternative expressions for the significance of the peak
observed in  Fig.~2 were considered. The first expression is the
na{\"i}ve estimator $\nsts/\sqrt{\nbts}$ used in
Refs.~\cite{SPring8,DIANA,JLab,SAPHIR,Asr03,JLabp}.
The corresponding result is listed in Table~\ref{tab1}. Because this
statistic neglects the uncertainty in the background fit, it
overestimates the significance of the peak~\cite{Ead71}. A
second estimator that was used in the analysis presented in
Ref.~\cite{Alt03}, $\nsts/\sqrt{\nsts+\nbts}$, gives a somewhat
lower value, but may still underestimate the background
uncertainty.
A third  estimate of the significance
is given by $N_{s}/{\delta N_{s}}$, where $N_s$ is now the full area
of the peak from the fit and $\delta
N_s$ is its fully correlated uncertainty. This ratio measures how
far the peak is away from zero in units of its own standard
deviation. All correlated uncertainties from the fit, including
those of the background parameters, are accounted for in $\delta
N_s$. The results obtained with this expression are also given in
Table~\ref{tab1}.

\begin{table*}[t]
\caption{ Masses and widths of observed invariant mass
peaks for four known particles, compared to the known
masses~\protect\cite{PDG} and the widths obtained from a
Monte Carlo simulation of the spectrometer.  (The widths
for the $\Xi^-(1321)$ are from mass-difference spectra.)
The $\Lambda(1520)$ peak of Fig.~3 was fitted with a Gaussian
folded with a Breit-Wigner form whose $\Gamma$ width was fixed
at the PDG value (15.6\,MeV). The uncertainties in the resulting mass and Gaussian
$\sigma$ width were inflated by the factor $\sqrt{\chi^2/N_{dof}}$.
In the last row, $P_{cm}$ is the momentum of each decay product
in the rest frame of the decaying particle.
}
\label{tab:masscal}
\begin{ruledtabular}
\begin{tabular}{lcccc}
& $K_S^0\rightarrow \pi^+\pi^-$ & $\Lambda(1116)\rightarrow p\pi^-$ &
$\Lambda(1520)\rightarrow pK^-$ & $\Xi^-(1321)\rightarrow p\pi^-\pi^-$\\ \hline
Observed mass [MeV] & $496.8\pm0.2$ & $1115.70\pm0.01$ & $1522.7\pm1.9$
& $1321.5\pm0.3$ \\
PDG Mass [MeV] & $497.67$ & $1115.68$ & $1519.5\pm1.0$ & $1321.31\pm0.13$ \\
$\sigma$ Width (data) [MeV] & $6.2\pm0.2$ & $2.6\pm0.1$
& $4.4\pm3.7$& $3.1\pm0.3$
\\ $\sigma$ Width (MC) [MeV] & $5.4$ & $2.1$
& $3.5$ & $2.5$
\\ Decay $P_{cm}$ [MeV/c] & 206 & 101 & 244 & 139 ($\Lambda\pi^-$)
\end{tabular}
\end{ruledtabular}
\end{table*}
The systematic uncertainty of the mass of the state observed in Fig.~2
is estimated to be $\pm 2.1$\,MeV by adding in quadrature
a contribution of $1$\,MeV to
account for the effect of using different spectrum analysis methods
(cf. Table~\ref{tab1}) plus a contribution of 1.9\,MeV from the
precision with which the spectrometer can reproduce known particle masses
(cf. Table~\ref{tab:masscal}).
The 1.9\,MeV contribution to the systematic uncertainty accounts for both
the discrepancies from the PDG~\cite{PDG} mass values and the
statistical precision of their fits. As an example, the
$\Lambda(1520)$ mass peak fitted with a Gaussian with free width convoluted with a 
Breit-Wigner form with its width fixed at the PDG value~\cite{PDG} is shown in the
$M_{pK^-}$ spectrum of Fig.~3.  The event selection for the spectra in
this figure is the same as for the $p K^0_S$ analysis, except the
reconstructed $K^0_S$ track is replaced by that of the observed charged
kaon, which is required to have a minimum momentum of 3\,GeV.
\begin{figure}[b] 
\begin{center}
\includegraphics[width=7.0cm]{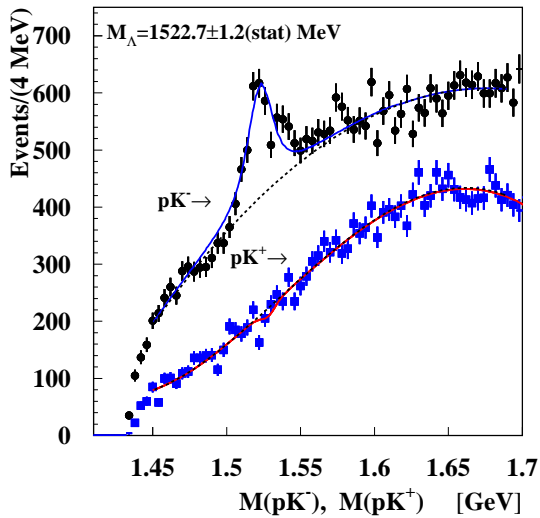}
\end{center}
\vspace{-5mm}

\caption{Spectra of invariant mass $M_{pK^-}$ (top) and
$M_{pK^+}$ (bottom). A clear peak is seen for the $\Lambda(1520)$
in the $M_{pK^-}$ invariant mass distribution. However, no peak
structure is seen for the hypothetical $\Theta^{++}$ in the
$M_{pK^+}$ invariant mass distribution near 1.53\,GeV.}

\label{fig3}
\end{figure}

The mass values reported to date for the $\Theta^+$ state by other
experiments are compared to the present results in Fig.~4, and listed
in Table~\ref{tab2} . The systematic uncertainties of the DIANA and
ITEP measurements were taken to be $\pm3$\,MeV in the absence of
explicit values quoted in the corresponding papers~\cite{DIANA,Asr03}.
By fitting the data with a
constant, a reduced $\chi^2$ value of $12.41/6$ is found, corresponding
to a confidence level of 0.053 as defined in the PDG~\cite{PDG}.
The weighted average~\cite{PDG} of the masses observed in all
experiments is $1536.2\pm2.6$\,MeV, which is represented by the shaded
band in Fig.~4. In evaluating this average mass value, the quadratic
sum of the statistical and systematic uncertainties of all
measurements are taken into account.
The uncertainty of the average was scaled by the usual factor
of square root of the reduced $\chi^2$.

\begin{figure} [b] 
\begin{center}
\includegraphics[width=7.0cm]{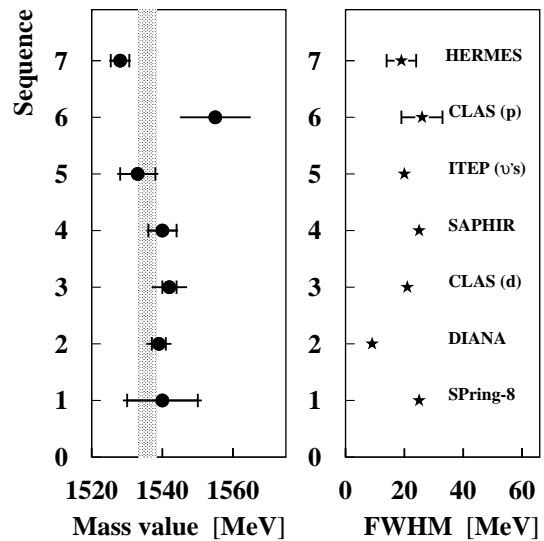}
\end{center}

\caption{Mass values and experimental FWHM widths observed in various
experiments for the $\Theta^+$ state. The inner error bars represent
the statistical uncertainties, and the outer error bars represent the
quadratic sum of the statistical and systematic uncertainties.
(Some uncertainties for the widths are not available from the
other experiments.)
The hatched area corresponds to the weighted average of all data $\pm1$
standard deviation.}

\label{fig4}
\end{figure}

Since no realistic model for the photoproduction of exotic baryons at
this experiment's energy is presently available, a ``toy Monte Carlo''
was produced to study the constraints imposed on the decay products by
only kinematics and acceptance. It generates parent particles with
specified mass and width, at vertices distributed according to the
HERMES target gas profile. The generated events were then passed to a
full simulation of the spectrometer that included the performance of
the RICH. The unknown kinematic distribution of the parent in
transverse momentum $P_t$ was taken to be Gaussian with a width of
$\sigma=0.4\,$GeV, which is typical of intrinsic transverse momentum
of partons or transverse momentum induced by the fragmentation
process, and also corresponds to the root-mean-square value of the
transverse momentum distribution of $\Lambda(1116)$ and
$\bar{\Lambda}$ hyperons that are observed at HERMES. The simulated
acceptance was found to be fairly insensitive to this parameter. The
longitudinal momentum $P_z$ was assigned a monotonically falling
distribution similar to what has been observed for $\Lambda$ hyperons
at HERMES.  The acceptance for the $p K^\pm$ final state is
insensitive to this choice, but this is not the case for the $p \pi^+
\pi^-$ final state, where drastically different assumptions can result
in a factor of two change in the acceptance.  The full width $\Gamma$
of the $\Theta^+$ was chosen to be 2\,MeV, according to the limit
recently derived from a review of $KN$ phase shift
analyses~\cite{Arn03}. The effect of such an intrinsic width is small
compared to that of the spectrometer resolution,
which was found in this simulation to be $\sigma=6.2$\,MeV
in $M_{p\pi^+\pi^-}$, or 4.3\,MeV in $M'_{pK^0_S}$ of Eq.~\ref{eq:massdiff}.
These values are assigned
a systematic uncertainty of $\pm1$\,MeV, based on the level of
agreement between observed and simulated widths of four known
particles, as shown in Table~\ref{tab:masscal}.

The width of the peak of interest in Fig.~2, given in Table~\ref{tab1}, is
somewhat larger than the instrumental resolution derived from the simulation.
An attempt was made to repeat the fits of Table~\ref{tab1} using for the
peak of interest a Breit-Wigner form convoluted with a Gaussian
whose width was fixed at the simulated resolution.\footnote{The software
tool RooVoigtian was used, for unbinned fitting.}
The resulting mass values are consistent with those given in Table~\ref{tab1},
and the resulting values for the intrinsic width are
 $\Gamma = 12\pm9$(stat)$\pm3$(syst)\,MeV in case a) of Table~\ref{tab1},
 $\Gamma = 20\pm8$(stat)$\pm3$(syst)\,MeV in case a'),
 $\Gamma = 8\pm8$(stat)$\pm3$(syst)\,MeV in case b), and
 $\Gamma = 13\pm9$(stat)$\pm3$(syst)\,MeV in case b').
The systematic uncertainties here correspond only to the $\delta\sigma=\pm1$\,MeV
uncertainty in the instrumental resolution, which was discussed above.

\begin{table} 
\caption{Mass values and experimental widths for the $\Theta^+$ state
as observed in the various experiments. The present result is also
listed. In calculating the weighted average of the data, the systematic
uncertainties of DIANA and ITEP are taken to be $\pm$3\,MeV.}
\label{tab2}
\begin{ruledtabular}
\begin{tabular}{lllc} Experiment & $\Theta^+$ mass  & FWHM &
Ref.  \\
 & (MeV)& (MeV)&  \\
 \hline
SPring8 & $1540\pm10\pm5$      & $25$ & \cite{SPring8} \\
DIANA   & $1539\pm2\pm$ ``few''  & $9 $ & \cite{DIANA} \\
CLAS (d) & $1542\pm2\pm5$       & $21$ & \cite{JLab} \\
SAPHIR  & $1540\pm4\pm2 $      & $25$ & \cite{SAPHIR} \\
ITEP ($\nu$'s) & $1533\pm5$ & $20$ & \cite{Asr03} \\
CLAS (p) & $1555\pm1\pm10$       & $26\pm7$ & \cite{JLabp} \\
HERMES  & $1528\pm2.6\pm2.1$ & $19\pm5\pm2$ & \\
\hline
world average & $1536.2\pm 2.6$ \\
\end{tabular}
\end{ruledtabular}
\end{table}

In view of the speculation that the observed resonance is
isotensor~\cite{Cap03}, the possibility that the $\Theta^{++}$
partner is present in the $M_{p K^+}$ spectrum was explored.
Although Fig.~3 shows a clear peak for the $\Lambda(1520)$ in the
$M_{pK^-}$ invariant mass spectrum, there is no peak structure
observed in the $M_{pK^+}$ invariant mass distribution. From a fit
(curve in Fig.~3) of the $M_{pK^+}$ distribution using a free
polynomial plus a Gaussian with the fixed location ($\pm5$\,MeV)
from the fit in Fig.~2b and a fixed width corresponding to
the simulated peak width of $\sigma=4.5$\,MeV
in this decay channel, the Gaussian area for a hypothetical $\Theta^{++}$
peak is found to be $-40\pm 30$ events.  This result is
robust against varying the order of the polynomial background.
It corresponds to an upper limit of zero counts at the 91\%
confidence level.

The failure to observe a $\Theta^{++}$ suggests that the $\Theta$
is likely to be isoscalar.  However, in the situation more
probable at lower beam energy that the $\Theta$ is produced only
via the exclusive reaction $\gamma + p \rightarrow \Theta + K$
without any other hadrons in the final state, the following
limitations would apply to deductions about its isospin.  Under
the assumption of isospin symmetry, selection rules limit the
transition amplitude for forming a tensor $\Theta$ to a single
reduced matrix element for an isovector transition.  In this
case, production of the $\Theta^{++}$ and the $\Theta^+$ are
expected to have comparable strength, and the failure to observe
the $\Theta^{++}$ rules out the $I=2$ assignment.  Production of an
isovector $\Theta$ would arise from a sum of three reduced amplitudes
with unknown magnitudes and phases.  With only model-dependent
values for these amplitudes, no precise statement can be made
about the relative yields of the $\Theta^{++}$ and $\Theta^+$. 
However, because a nearly complete cancellation is improbable,
the failure to observe the $\Theta^{++}$ indicates that an
isovector $\Theta$ is unlikely.

Estimates of the spectrometer acceptance times efficiency from the toy
Monte Carlo simulation can be used to estimate some cross sections.
Using the assumptions about the initial
kinematic distribution described above and assuming that the decay angle distribution
is flat, these acceptances were
estimated to be 1.5\% for both $\Lambda(1520)\rightarrow p K^-$ and
$\Theta^{++}\rightarrow p K^+$, and 0.05\% for $\Theta^+\rightarrow p
K^0_S$.
Taking the branching fraction of the $\Theta^+$ to $p K^0_S$ to be $(1/2)\cdot(1/2)$
(to account for competition with both the $n K^+$ channel and $K^0_L$),
the cross section for its photoproduction is
found to range from about 100 to 220\,nb $\pm 25\%\mbox{(stat)}$, depending
on the model for the background and the functional form fitted to the peak. 
The cross section for photoproduction
of the $\Lambda(1520)$ is found to be $62\pm11\mbox{(stat)}$\,nb.  Hence the
ratio of the $\Theta^{+}$ cross section to that for the
$\Lambda(1520)$ is found to lie between 1.6 and 3.5. All of these estimates
are subject to an additional factor of two uncertainty, to account for the
assumptions about the kinematic distribution of the parents used in
the simulation as explained above, and neglected trigger inefficiencies.

In conclusion, evidence has been obtained in quasi-real
photoproduction on a deuterium target for a narrow baryon state
in the $pK^{0}_{S}$~invariant mass spectrum at
$1528\pm2.6\mbox{(stat)}\pm2.1\mbox{(syst)}$\,MeV. 
Depending on the background model, the width of the observed peak
may be larger than the experimental resolution of
$\sigma=$4.3--6.2\,MeV. 
Fitting the peak with a convolution of a Breit-Wigner shape with a
Gaussian representing the simulated instrumental resolution yields 
an extracted intrinsic width 
$\Gamma = 17\pm9$(stat)$\pm3$(syst)\,MeV (the average of the results
from cases a' and b' of Table~\ref{tab1}).
The significance of the observed state
expressed as $\nsts/\sqrt{\nbts}$ ranges from $4.2$\,$\sigma$ to
$6.3$\,$\sigma$, and expressed as $N_s/\delta N_s$ ranges from
$3.4$\,$\sigma$ to $4.3$\,$\sigma$, again depending on the model for
the background. This observation provides further evidence for
the existence of a narrow baryon state with $|S|=1$ and a mass in
the region where such a feature was observed by earlier
experiments.  Formally, the difference between the value for the
mass derived here and that from the other experiments reduces the
confidence level of the combined fit to all mass data from 0.57 to
0.053, a value still typical for well-established
particles~\cite{PDG}. There is no identified $\Sigma^{*+}$ state
with $S$=$-1$ in the invariant mass region between 1500 and
1550\,MeV. Therefore, the state observed here may be interpreted
as the predicted exotic $\Theta^+$ pentaquark $S$=$+1$ baryon.
The absence of a corresponding signal in the $pK^+$ invariant
mass spectrum indicates that the observed $\Theta^+$ is not isotensor
and is probably an isosinglet.

\begin{acknowledgments}

We thank Markus Diehl, T.-S.H. Lee, Kim Maltman, Maxim Polyakov and 
Craig Roberts for stimulating
discussions. We gratefully acknowledge the DESY management for its
support and the staff at DESY and the collaborating institutions for
their significant effort. This work was supported by the FWO-Flanders,
Belgium; the Natural Sciences and Engineering Research Council of
Canada; the National Natural Science Foundation of China; the INTAS
and ESOP network contributions from the European Community; the German
Bundesministerium f\"ur Bildung und Forschung; the Deutsche
Forschungsgemeinschaft (DFG); the Deutscher Akademischer
Austauschdienst (DAAD); the Italian Istituto Nazionale di Fisica
Nucleare (INFN); Monbusho International Scientific Research Program,
JSPS, and Toray Science Foundation of Japan; the Dutch Foundation for
Fundamenteel Onderzoek der Materie (FOM); the U. K. Engineering and
Physical Sciences Research Council and the Particle Physics and
Astronomy Research Council; and the U. S. Department of Energy and the
National Science Foundation.

\end{acknowledgments}

\end{document}